\documentclass[preprint,12pt]{elsarticle}
\usepackage{CJK}
\usepackage{amsmath}
\usepackage{amssymb}
\usepackage[pdftex,colorlinks]{hyperref}
\usepackage{multirow}

\biboptions{compress}

\begin{document}

\begin{frontmatter}

\title{Investigating the linear structure of Boolean functions based on Simon's period-finding quantum algorithm}

\author{Li Yang$^1$\corref{1}}%\ead{yangli@iie.ac.cn}
\author{Hong-Wei Li$^{1,2}$}
\cortext[1]{Corresponding author. E-mail: yangli@iie.ac.cn}
\address{1.State Key Laboratory of Information Security, Institute of Information Engineering, Chinese Academy of Sciences, Beijing 100093, China\\
2.Mathematics Department, Henan Institute of Education, Zhengzhou, 450046, China}
\begin{abstract}
It is believed that there is no efficient classical algorithm to determine the linear structure of Boolean function. We investigate an extension of Simon's period-finding quantum algorithm, and propose an efficient quantum algorithm to determine the linear structure of Boolean function.

%We propose a quantum algorithm to determine the linear structure of a Boolean function executed by a quantum oracle, and show that Simon's original problem can also be solved with this method.
%For a Boolean function provided with polynomial size representation, we find an efficient algorithm to compute its linear structure.
%In addition, we present some interesting properties of the linear structure of Boolean functions.

\end{abstract}


\end{frontmatter}

% \linenumbers


D.R.Simon's paper \cite{DR97} work on the comparison between two algorithms based on the access to a classical or quantum oracle of the Boolean function considered, respectively. He suggest a quantum algorithm to distinguish between two classes of functions, one is one-to-one, another has a period. Simon's result means that in the circumstance with a quantum oracle executing a $2\rightarrow1$ periodic multiple Boolean function, one can find the period efficiently. Here we generalize this algorithm to determine the linear structure of Boolean function. Understanding the linear structure of a Boolean function is important while designing a cryptographic algorithm or executing some cryptanalysis. The cryptologists have presented various relevant results, such as that in \cite{SD01,EC97,FX95,CZX11}, though there is still no polynomial algorithm suggested. Here we provide a polynomial size quantum algorithm to determine the linear structure of Boolean functions via extending Simon's period-finding algorithm.

\section{Using Simon's algorithm to find multiple periods}

Given a multi-output Boolean function $F:\{0,1\}^n\rightarrow\{0,1\}^{n-1}$,  if there is a nontrivial string $s$ of length $n$ such that $\forall x, F(x)=F(x\oplus s)$
($\oplus$ always denote bitwise exclusive-or in this note), $s$ will be called the period of $F(x)$.
Simon's quantum algorithm gave a solution for finding $s$.

The Simon's algorithm is composed of $O(n)$ repetitions of the following routine.

1.\;Perform the Hadamard transformation $H^{(n)}$ on a qubit-string in state $|0\cdots 0\rangle$, change it into state $2^{-n/2}\sum_x|x\rangle_I$.

2.\;Compute $F(x)$, to get $2^{-n/2}\sum_x|x\rangle_I|F(x)\rangle_{II}$.

3.\;Perform $H^{(n)}$ on the register $I$, Producing $2^{-n}\sum_y\sum_x(-1)^{x\cdot y}|y\rangle_I|F(x)\rangle_{II}$.

Suppose there is some $s\neq 0$, such that $\forall x, F(x)=F(x\oplus s)$. Then for each $y$, $|y, F(x)\rangle=|y, F(x\oplus s)\rangle$,
and the amplitude of this configuration will be
\begin{equation}\alpha(x,y)=2^{-n}((-1)^{x\cdot y}+(-1)^{(x\oplus s)\cdot y})
=2^{-n}(-1)^{x\cdot y}[1+(-1)^{s\cdot y}] \end{equation}
Measure the first register, getting a value of $y$, which must satisfies $s\cdot y=0$.

 After $O(n)$ repetitions of the routine, $n$ linearly independent values of $y$ can be
 obtained. Then $s$ can be obtained by solving the linear system of equations
 \begin{equation}\label{eq:f}
 \left\{\begin{array}{ll}
 y_1\cdot s=0\\y_2\cdot s=0\\\cdots\\y_n\cdot s=0.
 \end{array}\right.
 \end{equation}

If the set of Equs. (\ref{eq:f}) has only one nontrivial solution, that is
Simon's original quantum algorithm, but if it has more than one solutions, the solutions are all the periods of $F$. In other word, Simon's quantum algorithm can also be applied to the situation that  Boolean function $F$ of output $n$ has two or more periods.

Suppose $\{b_1,b_2,\cdots\cdots b_k\}$ are the linear independent periods of $F$.

Perform the quantum oracle to compute $F(m)$, get
$$2^{-n/2}\sum_{m=0}^{2^n-1}|m\rangle_I|F(m)\rangle_{II}.$$
Measure the second register, which leads the first register collapse into the state:
$$\frac{1}{\sqrt{2^k}}\sum_{(\alpha_1,\cdots,\alpha_k)=0}^{2^k-1}|m\oplus\alpha_1b_1\oplus\cdots\oplus\alpha_kb_k\rangle_I.$$

Then, perform Hadamard transform on the first register:
\begin{equation}
H^{(n)}\left[\frac{1}{\sqrt{2^k}}\sum_{(\alpha_1,\cdots,\alpha_k)=0}^{2^k-1}|m\oplus\alpha_1b_1
\oplus\cdots\oplus\alpha_kb_k\rangle_I\right]\nonumber
\end{equation}
\begin{equation}\label{eq:a}
=\frac{1}{\sqrt{2^k}}\frac{1}{\sqrt{2^n}}
\sum_y(-1)^{m\cdot y}[1+(-1)^{b_1\cdot y}]\cdots[1+(-1)^{b_k\cdot y}]|y\rangle_I,
\end{equation}
and measure the register, we obtain the vectors $y^{(1)},\cdots,y^{(m)}\in F_2^n$. It is necessary that
 \begin{equation}\left\{\begin{array}{ll}
 b_1\cdot y^{(1)}=0,\\\vdots\\b_k\cdot y^{(1)}=0,
 \end{array}\right.,\quad\cdots,
 \quad
 \left\{\begin{array}{ll}
 b_1\cdot y^{(m)}=0,\\\vdots\\b_k\cdot y^{(m)}=0.
 \end{array}\right.
 \end{equation}

There are $m$ blocks (totally $m\times k$  equations) above. We can divide them into $k$ groups to obtain $b_1,\cdots,b_k$:
 \begin{equation}\label{eq:b}\left\{\begin{array}{ll}
 y^{(1)}\cdot b_1=0,\\\vdots\\y^{(m)}\cdot b_1=0,
 \end{array}\right.,\quad\cdots,
 \quad
 \left\{\begin{array}{ll}
y^{(1)}\cdot b_k=0,\\\vdots\\y^{(m)}\cdot b_k=0.
 \end{array}\right.
 \end{equation}
Equs.(\ref{eq:b}) is actually the same as Simon's original Equs. (\ref{eq:f}).
\section{The quantum algorithm for finding the linear structure of Boolean function}
\noindent{\bf Definition 1}\quad
Let $f(x):F^n_2\rightarrow F_2$ is an Boolean function.
Suppose $\alpha \in F^n_2$.  If $\forall x \in F^n_2, f(x\oplus\alpha)+f(x)=c=f(\alpha)+f(0)$, we call $\alpha$ is a linear structure of $f(x)$.

Let $U_f$ denote the collection of the linear structure of $f(x)$.
$$U_f^{(0)}=\{\alpha \in F^n_2|f(x\oplus\alpha)+f(x)=0, \forall x \in F^n_2\}.$$
$$U_f^{(1)}=\{\alpha \in F^n_2|f(x\oplus\alpha)+f(x)=1, \forall x \in F^n_2\}.$$
Obviously $U_f=U_f^{(0)}\bigcup U_f^{(1)}.$ Inspired by the Simon's period-finding method,
we can use the following quantum algorithm to find out a  basis of $U_f^{(0)}$:\\
\noindent{\bf Algorithm}

Initially $i=1$, $r=n$.

(1) Randomly choose an integer $l_i$ (we choose $l_{i}\geq n$ and $l_{i+1}> l_{i}$), and generate a set of random vectors $a_1^{(i)},\cdots,a_{l_i}^{(i)}\in F^n_2$, and then compute
 $$\frac{1}{\sqrt{2^n}}\sum_{m=0}^{2^n-1}|m\rangle_I|0\rangle_{II}\rightarrow \frac{1}{\sqrt{2^n}}\sum_{m=0}^{2^n-1}|m\rangle_I
 |f(m),f(m\oplus a_1^{(i)}),\cdots f(m\oplus a_{l_i}^{(i)})\rangle_{II}.$$

(2) Measure the register $II$,  suppose the output is $(F_0,\cdots,F_{l_i})$, then the quantum register $I$ collapse to (let $b_0^{(i)}=0$)
\begin{equation}\label{eq:n}|\psi_f\rangle_I=\frac{1}{\sqrt{N_i+1}}\sum_{j=0}^{N_i}|m\oplus b_j^{(i)}\rangle_I,\end{equation}
where $m, m\oplus b_1^{(i)},\cdots, m\oplus b_{N_i}^{(i)}$ satisfy the following equations:
\begin{equation}\left\{\begin{array}{ll}
f(x)=F_0,\\\vdots\\f(x\oplus a_{l_i}^{(i)})=F_{l_i}.
 \end{array}\right.
 \end{equation}
Begin with (\ref{eq:n}), by the method of Section 1, perform the Hadamard transformation on the register $I$, and measure the register, repeat the experiment $An$(where $A$ is a constant) times, we will get a group of linearly independent $y$, solve a linear system of equations like Equs.(\ref{eq:b}), we will get a solution basis $b_1^{'(i)},\cdots,  b_{N_i}^{'(i)}$.
Let $\mathfrak{L}(b_1^{'(i)},\cdots,  b_{N_i}^{'(i)})$ denote the space generated by the vector $b_1^{'(i)},\cdots,  b_{N_i}^{'(i)}$. By the definition of $U_f^{(0)}$ we know that there must be
$U_f^{(0)}\subseteq \mathfrak{L}(b_1^{(i)},\cdots,  b_{N_i}^{(i)})$,
but it is not necessary that
$\mathfrak{L}(b_1^{(i)},\cdots,  b_{N_i}^{(i)})\subseteq U_f^{(0)}$.
So after we solve the answers $b_1^{'(i)},\cdots,  b_{N_i}^{'(i)}$, they may not be the basis of $U_f^{(0)}$. For deal with that situation, we
do the following steps:

(3) If $i<r$, take $i=i+1$, repeat the steps $(1)$ and $(2)$, and output
a set of solutions $b_1^{'(i+1)},\cdots,  b_{N_{i+1}}^{'(i+1)}$.
Now we have obtained $r$ sets of solutions
\begin{equation}\begin{array}{ll}
b_1^{'(1)},\cdots,  b_{N_1}^{'(1)};\\\quad\vdots\\b_1^{'(r)},\cdots,  b_{N_r}^{'(r)}.
 \end{array}
 \end{equation}
If $\exists i_0 < r$ satisfies: for any $i,i'\geq i_0, i\neq i'$,
$\mathfrak{L}(b_1^{'(i)},\cdots,  b_{N_i}^{'(i)})=\mathfrak{L}(b_1^{'(i')},\cdots,  b_{N_{i^{'}}}^{'(i')})$,
output $b_1^{'(r)},\cdots,  b_{N_r}^{'(r)}$, and go to (4); otherwise let $r=r+1$, and go to (3).

(4) Test and verify: after we get the output $b_1^{'(r)},\cdots,  b_{l_r}^{'(r)}$,
take $l_r$ vectors into the representation of $f(x)$ separately. That is
choose $p(n)$ values of $x$, and compute whether it is $f(x)=f(x\oplus b_i^{'(r)})|_{i=1,\cdots,l_r}$. If we don't find any value of $x$ violates the equation, then we will confirm $b_1^{'(r)},\cdots,  b_{l_r}^{'(r)}$ is a basis of $U_f^{(0)}$.

\section{The Simplified quantum algorithm}
\noindent{\bf Algorithm}

(1) Generate a set of linear independent vectors $a_1,\cdots,a_{n}\in F^n_2$, and then compute
 $$\frac{1}{\sqrt{2^n}}\sum_{m=0}^{2^n-1}|m\rangle_I|0\rangle_{II}\rightarrow \frac{1}{\sqrt{2^n}}\sum_{m=0}^{2^n-1}|m\rangle_I
 |f(m),f(m\oplus a_1),\cdots f(m\oplus a_{n})\rangle_{II}.$$

 Measure the register $II$,  suppose the output is $(F_0,\cdots,F_{n})$, then the quantum register $I$ collapse to (let $b_0'=0$)
\begin{eqnarray} \label{eq:j}
|\psi_f\rangle_I=\frac{1}{\sqrt{N+1}}\sum_{j=0}^{N}|m\oplus b_j'\rangle_I,\end{eqnarray}
where $m, m\oplus b_1',\cdots, m\oplus b_N'$ satisfy the following equations:
\begin{equation}\left\{\begin{array}{ll}
f(x)=F_0,\\\vdots\\f(x\oplus a_{n})=F_{n}.
 \end{array}\right.
 \end{equation}
By the definition of $U_f^{(0)}$ we know that there must be
$U_f^{(0)}\subseteq \{b_0',\cdots,  b_N'\}$,
but it is not necessary that
$\{b_1',\cdots, b_N'\}\subseteq U_f^{(0)}$.
Perform the Hadamard transformation on the register $I$, and by the Equ. (\ref{eq:j})
we get
\begin{eqnarray} \label{eq:k}
H^{(n)}\left[\frac{1}{\sqrt{N+1}}\sum_{j=0}^{N}|m\oplus b_j'\rangle_I\right]
=\frac{1}{\sqrt{N+1}}\frac{1}{\sqrt{2^n}}\sum_y\sum_{j=0}^{N}
(-1)^{(m\oplus b_j')\cdot y}|y\rangle_I.
\end{eqnarray}
Measure the register to get $y$.

(2) After $\alpha(n)$ (polynomial function of $n$, refer to the analysis of the algorithm)  times repeat of (1), we have obtained
\begin{equation}\begin{array}{ll}
y_1,\cdots,  y_{j}.
 \end{array}
 \end{equation}

(3) Solve the linear system of equations generalized by $y_1,\cdots,  y_{j}$ the same as  Equs. (\ref{eq:f}) to get the solution $b_1,\cdots,  b_{n-j}$.

(4) Test and verify: after we get the output $b_1,\cdots,  b_{n-j}$,
take $l$ vectors into the representation of $f(x)$ separately. That is
choose $p(n)$ values of $x$, and compute whether it is $f(x)=f(x\oplus b_i)|_{i=1,\cdots,n-j}$. If we don't find any value of $x$ violates the equation, then we will confirm $b_1,\cdots,  b_{n-j}$ is a basis of $U_f^{(0)}$.

\noindent{\bf The analysis of the algorithm}

Let $ F^n_2$ be a space which is composed of $\{0,1\}$ strings of length $n$. Suppose the probability distribution over $ F^n_2$ is uniform, we randomly choose the elements of $ F^n_2$. Denote the event of getting $n$ linearly independent elements through $k(k\geq n)$ times picking by $A_k$. $P(A_k)$ is the probability of the event $A_k$ happens. If $A_k$ happens, we will say that it is successful. Otherwise, we say that it is failed. Then
\begin{eqnarray}
P(A_n)=(1-\frac{1}{2^n})(1-\frac{1}{2^{n-1}})\cdots (1-\frac{1}{2})=P_n.
\end{eqnarray}
If $k>n$,
\begin{eqnarray}
P(A_k)=P_n\sum_{x_0+x_1\cdots +x_{n-1}+x_n=k-n}\frac{1}{2^{nx_0+(n-1)x_1+\cdots +x_{n-1}}}.
\end{eqnarray}
Let $i=k-n$,
\begin{eqnarray}q(n,i)=\sum_{x_0+x_1\cdots +x_{n-1}+x_n=i}\frac{1}{2^{nx_0+(n-1)x_1+\cdots +x_{n-1}}},\end{eqnarray}
then
\begin{eqnarray}\label{eq:l}q(n,i)&=&\sum_{x_0+x_1\cdots +x_{n-1}=0}^{i}\frac{1}{2^{nx_0+(n-1)x_1+\cdots +x_{n-1}}}
\nonumber\\
&=&q(n-1,0)+\frac{1}{2}q(n-1,1)+\cdots +\frac{1}{2^i}q(n-1,i)
\nonumber\\
&=&\sum_{m=0}^i(1/2^m)\cdot q(n-1, m).
\end{eqnarray}
\begin{eqnarray}\label{eq:m}q(1,i)=\sum_{x_0+x_1=i}\frac{1}{2^{x_0}}=\sum_{x_0=0}^i\frac{1}{2^{x_0}}=2-\frac{1}{2^i}.
\end{eqnarray}
Let $s(n,k)=P(A_{k})=P_n\cdot q(n,k-n)$, $h(n,k)=\log(1-s(n,k)).$
By means of Mathematica and with the application of (\ref{eq:l}), (\ref{eq:m}), we draw the function image of $s(n,k)$ and $ h(n,k)$.

\begin{figure}
\begin{center}
\begin{minipage}[c]{0.55\textwidth}
\centering
\includegraphics[angle=0,width=5cm,height=5cm]{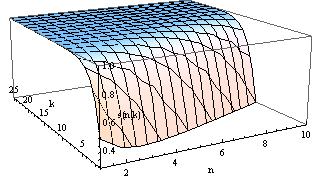}
\renewcommand{\figurename}{fig}
\centering
\caption{The successful probability of picking  \;\;\;\;\;\;\;\; k times from an n dimensional space}
%\label{fig:levfig}
\end{minipage}%
\begin{minipage}[c]{0.55\textwidth}
\centering
\includegraphics[angle=0,width=5cm,height=5cm]{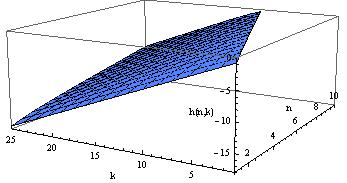}
\renewcommand{\figurename}{fig}
\caption{The logarithm of the failed probability of picking k times from an n dimensional space}
%\label{fig:levfig}
\end{minipage}
\end{center}
\end{figure}

%\begin{figure}[htb]
 %\centering
%\includegraphics[scale=0.4,bb= 200 0 695 260]{pic-1.jpg}\\
 % \caption{Relations between probabilities the adversary get and the error rate caused by the attacks}\label{pic1}
%\end{figure}

%\begin{figure}[htb]
% \centering
%\includegraphics[scale=0.4,bb= 200 0 695 260]{pic-2.jpg}\\
%  \caption{$h(n,k)$}\label{pic2}
%\end{figure}

In conclusion, if $f(x)$ has a linear structure, we can find it in the time of linear function of $n$.  If $f(x)$ has no linear structure, see below.

\noindent{\bf Definition 2}\quad
Let $f(x):F^n_2\rightarrow F_2$ be a Boolean function.
Suppose $\alpha \in F^n_2$. Given an integer $r$, $0\leq r\leq 2^n$, If $\exists$ a set $V(\alpha): |V(\alpha)|\leq r$, such that
$\forall x \in F^n_2\setminus V(\alpha), f(x\oplus\alpha)+f(x)=c=f(\alpha)+f(0)$, we call $\alpha$ is a $r$-type linear structure of $f(x)$.
Specially, if $V(\alpha)$ is independent with the choice of $\alpha$, without of generality, $|V(\alpha)|=r$, we say $\alpha$ is a uniform $r$-type linear structure of $f(x)$. Obviously,
If $r=0$, the uniform $0$-type linear structure of $f(x)$ is the linear structure of $f(x)$ which is defined in Definition 1.

If $0<r\ll2^n$, $1-\frac{r}{2^n}\approx 1$. If $f(x)$ has a uniform $r$-type linear structure, we choose $l$ numbers of different $a_n$ in the experiment and fix it, and we do $p(n)$ times of experiments, we will obtain the erroneous conclusion that  $f(x)$ has a linear structure with a probability $(1-\frac{r}{2^n})^{(l+1)p(n)}$. If we want this probability no more than $\frac{1}{2^{\beta n}}$, we should let
$p(n)>-\frac{\beta n}{l+1}\log_{1-\frac{r}{2^n}}^2$. Because when $r<2^{n-1}$,
$\frac{\ln 2}{2\cdot \frac{r}{2^n}}<-\log_{1-\frac{r}{2^n}}^2=\frac{\ln 2}{-\ln (1-\frac{r}{2^n})}<\frac{\ln 2}{\frac{r}{2^n}}$.
So if we want obtain the fact that $f(x)$ has no linear structure, we should do the experiment $\frac{\beta n \ln 2}{l+1}\frac{2^n}{r}$ times, in other words, exponential times. If we do polynomial times of experiment, we will find a linear structure $\alpha$, which is not the true linear structure, but just a $r$-type one, we call it pseudo linear structure, but this pseudo linear structure is also helpful in the cryptanalysis.

\noindent{\bf Remark} In fact, solving whatever equations in the set of Equs. (\ref{eq:b}) can be accomplished by  quantum algorithm,
and quantum algorithm contains only the generation of the states, $n$ multiple Hadamard transform, and the measurement of the quantum states.
Specific processes are as follows.\\
(1)\quad The generation of the states: we can obtain the following $n$ qubit through  Hadamard transform of some particular qubit and the CNOT operation of
some qubit pair.
 \begin{eqnarray}|0\rangle_I&\rightarrow&\frac{1}{\sqrt{2}}\left(|0\rangle_I+|y^{(1)}\rangle_I\right)
\nonumber\\
&\rightarrow&
 \frac{1}{2}\left(|0\rangle_I+|y^{(1)}\rangle_I+|y^{(2)}\rangle_I+|y^{(1)}\oplus y^{(2)}\rangle_I\right)
 \nonumber\\
&\rightarrow&\cdots
\nonumber\\
&\rightarrow&
\frac{1}{\sqrt{2^m}}\sum_{(\alpha_1,\cdots,\alpha_m)=0}^{2^m-1}|\alpha_1y^{(1)}\oplus\cdots\oplus\alpha_my^{(m)}\rangle_I.
\end{eqnarray}
(2)\quad $n$ multiple Hadamard transform: perform $n$ multiple Hadamard transform on the above quantum state, producing
\begin{equation}\frac{1}{\sqrt{2^{m+n}}}\sum_z[1+(-1)^{y^{(1)}\cdot z}]\cdots[1+(-1)^{y^{(m)}\cdot z}]|z\rangle_I.\end{equation}
(3)\quad Measure the register $I$, and we can obtain some linear independent vectors $z^{(1)},\cdots,z^{(k)}$.
They are the solutions of Equs. (\ref{eq:b}): $b_1=z^{(1)},\cdots,b_k=z^{(k)}$.

\section{Discussion and conclusion}
Simon's algorithm is an important achievement in the history of quantum algorithm research, which inspired Shor's landmark work\cite{PW97}. Simon's paper investigates a fundamental theoretical problem  of computational complexity: whether a quantum computer can get an exponential acceleration compared to a classical computer, or, whether the strong Church-Turing thesis still holds under the environment of quantum computing. Since that IFP, DLP, et al. have not been proved without polynomial classical algorithms, the goal of Simon is still a challenging task.

Following Simon's idea, we propose an efficient quantum algorithm to determine the linear structure of a quantum oracle that executing a Boolean function, which probably be helpful in cryptographic designing and cryptanalysis.

\section*{Acknowledgement}

This work was supported by the National Natural Science Foundation of China under Grant No.61173157.

\noindent{\bf Appendix A. Some properties of the linear structure of Boolean function}\\

Suppose $f(x):F_2^n\rightarrow F_2$ is a Boolean function of $n$ variables.
If we know the algebraic normal form of $f(x)$, then we can ask whether there is a $s\in F_2^n$, such that $f(x\oplus s)=f(x)$?
 Obviously, if $s\in U_f^{(0)}$,  $f(x\oplus s)=f(x)$.
The following are some conclusions about how to find $U_f^{(0)}$ in classical algorithms.

Let
\begin{eqnarray} \label{eq:d}
f(x)&=&a_0+a_1x_1+a_2x_2+\cdots+a_nx_n+a_{12}x_1x_2
\nonumber\\
&&+\cdots+a_{n-1,n}x_{n-1}x_n+\cdots+a_{12\cdots n}x_1x_2\cdots x_n
\end{eqnarray}
where $x=(x_1,x_2,\cdots x_n)\in F_2^n, a_{i_1i_2\cdots i_r} \in F_2$, and $+$ is the sum mod 2.

Suppose $s=(s_1,s_2,\cdots s_n)\in U_f^{(0)}$,
\begin{eqnarray}\label{eq:g}
g(x)&\equiv&f(x\oplus s)+f(x)\nonumber\\
&=&a_1s_1+a_2s_2+\cdots+a_ns_n
+a_{12}(s_1x_2+s_2x_1+s_1s_2)\nonumber\\
&&+\cdots+a_{n-1,n}(s_{n-1}x_n+s_nx_{n-1}+s_{n-1}s_n)
+\cdots+a_{12\cdots n}s_1x_2\cdots x_n\nonumber\\
&&+\cdots+a_{12\cdots n}x_1x_2\cdots s_n+\cdots+a_{12\cdots n}s_1s_2x_3\cdots x_n
\nonumber\\
&&+\cdots+a_{12\cdots n}s_1s_2\cdots s_n.
\end{eqnarray}
We have
\begin{eqnarray}\label{eq:h}
\forall x, f(x)=f(x\oplus s)\Leftrightarrow g(x)=0.
 \end{eqnarray}

Using the following lemma, we find some relations between $U_f^{(0)}$ and the coefficients of $f(x)$
by exploiting the representation of $g(x)$.

 \noindent{\bf Lemma 1\cite{JY91}.}
Let $Z[x_1,\cdots,x_k]$ be a class of the polynomial with integer coefficient relative to uncertainty element
$x_1,\cdots,x_k$. $\forall T\subseteq I_k=\{1,2,\cdots,k\}$, $x_T=\prod_{i\in T}x_i$(Let $x_\phi=1$). Let
$Z_0[x_1,\cdots,x_k]$ denote the set of  all the elements in $Z[x_1,\cdots,x_k]$ which can be represent as the polynomials as following:
 \begin{eqnarray}
f(x_1,\cdots,x_k)=\sum_{T\subseteq I_k}b(T)x_T
 \end{eqnarray}
In other words, every item of $f(x_1,\cdots,x_k)$ is a product of some different variables with a integer coefficient.
If $f(x_1,\cdots,x_k)\in Z_0[x_1,\cdots,x_k]$, and $\forall a_i\in\{0,1\}(i=1,2,\cdots,k)$, we have
$f(a_1,\cdots,a_k)=0$, then $f(x_1,\cdots,x_k)\equiv0.$

Combine the above lemma and
(\ref{eq:h})
, we obtain the following result:

 \noindent{\bf Theorem 2.}
 { If $\exists a_{i_1i_2\cdots i_{r}}=1$,
$\forall k\geq r+1, a_{i_1i_2\cdots i_k}=0, $  Let $C_t=\{1,2,\cdots,n\}-\{i_1,i_2,\cdots,i_t\}$.
A sufficient and necessary  condition for a $s\in U_f^{(0)}$ is $g(x)\equiv 0$. Specifically, every coefficient of $g(x)$ is 0, i.e.
 \begin{eqnarray} \label{eq:c}
 \sum_{l\in C_t} a_{i_1,i_2,\cdots,i_{t},l}s_l+\cdots+
 \sum_{l_1,\cdots l_{r-t}\in C_t} a_{i_1,i_2,\cdots,i_{t},l_1,\cdots l_{r-t}}s_{l_1}\cdots s_{l_{r-t}}=0
 \end{eqnarray}
  for $\forall t, 0\leq t\leq r-1.$

\noindent{\bf Property 1.} {\it If $a_{12\cdots n}=1$, then it must be $s=0$, i.e. $U_f^{(0)}=\{0\}$. }\\
 \noindent{\bf Proof}  \qquad
In Equs. (\ref{eq:c}), if $t=r-1$, then
\begin{eqnarray} \label{eq:i}
 \sum_{l\in C_{r-1}} a_{i_1,i_2,\cdots,i_{r-1},l}s_l=0.
 \end{eqnarray}
specially, if $a_{12\cdots n}=1$, i.e. $r=n$, then (\ref{eq:i}) becomes
\begin{eqnarray}a_{12\cdots n}s_i=0\,(i=1,2,\cdots n).\nonumber
\end{eqnarray}

If $a_{12\cdots n}=1,$ then $s_1=s_2=\cdots=s_n=0$, that is $s=0$.

\noindent{\bf Property 2.} {\it If $a_{12\cdots n}=0$, there exist $a_{i_1i_2\cdots i_{n-1}}\neq 0$, there exist $s\in U_f^{(0)}$, $s\neq 0$,
then it must be $s=(s_1,s_2,\cdots s_n)=(a'_1,a'_2,\cdots a'_n) $, where $a'_i$ represents $a_{i_1i_2\cdots i_{n-1}}$ whose subscript doesn't have $i$.
Particularly, if $(a'_1,a'_2,\cdots a'_n)=(1,1,\cdots,1)$, then $s=(1,1,\cdots,1)$.}

 \noindent{\bf Proof}  \qquad Similar to Property 1,  if $a_{12\cdots n}=0$, there exist $a_{i_1i_2\cdots i_{n-1}}\neq 0$, then $r=n-1$, from (\ref{eq:i}) we get
 $$a_{23\cdots n}s_2+a_{13\cdots n}s_1=0.$$

 For more generally, $$a'_is_j+a'_js_i=0.$$
 If $a'_i=1, a'_j=0$, then $s_j=0=a'_j$. Since $s\neq 0$, we must have $s_i=1=a'_i.$

 If again we have $a'_k=1$, then $s_k+s_i=0$, so $s_k=s_i=1=a'_i=a'_k.$

 We then get $s=(s_1,s_2,\cdots s_n)=(a'_1,a'_2,\cdots a'_n). $

 \noindent{\bf Property 3.} {\it If $a_{12\cdots n}=0$,  $\forall a_{i_1i_2\cdots i_{n-1}}=0$, $\forall a_{i_1i_2\cdots i_{n-2}}=1$, $n\geq 4$,
 then it must be $s=0$, i.e. $U_f^{(0)}=\{0\}$. }

 \noindent{\bf Proof}  \qquad Similar to Property 1 and Property 2, and in this circumstance, $r=n-2$, by (\ref{eq:i}), we have
$$a'_{ij}s_k+a'_{jk}s_i+a'_{ik}s_j=0.$$
If $\forall a'_{ij}=1$, then $s_k+s_i+s_j=0$, so there is at least a 0 in  $s_k, s_i, s_j$, suppose $s_i=0$. And also
$$s_k+s_l+s_j=0,$$ we have $s_i+s_l=0$, so $s_l=s_i=0$. And that we have
$$s_i+s_l+s_j=0,$$ so $s_j=0$. And then $s_k=0$. So $s=0$.

\noindent{\bf Property 4.} {\it If $\forall a_{i_1i_2\cdots i_{n-2m}}=1$,
$\forall a_{i_1i_2\cdots i_{k}}=0, k\geq n-2m+1$, $n\geq 2m+2$,
 then it must be $s=0$, i.e. $U_f^{(0)}=\{0\}$. }

 \noindent{\bf Proof}  \qquad Similar to Property 3, and in this circumstance, $r=n-2m$, by (\ref{eq:i}), we first obtain
 $$\sum_{k=1}^{2m+1}s_{i_k}=0.$$
There is at least one  $s_{i_k}$ satisfies $s_{i_k}=0$. And also another $s_{i_{2m+2}}=0$. And then
the sum of the $s_{i_k}$ of $2m-1$ numbers is 0. Subsequently the sum of $2m-3,\cdots,3$ numbers is 0.
So every $s_i=0$, that is $s=0$.

\noindent{\bf Property 5.} {\it If $\forall a_{i_1i_2\cdots i_{n-2m+1}}=1$,
$\forall a_{i_1i_2\cdots i_{k}}=0, k\geq n-2m+2$, $n\geq 2m+1$,
there exist $s\in U_f^{(0)}$, $s\neq 0$,
then it must be $s=(1,1,\cdots,1)$. }

 \noindent{\bf Proof}  \qquad Similar to Property 4, and in this circumstance, $r=n-2m+1$,  by (\ref{eq:i}), we first obtain
 $$\sum_{k=1}^{2m}s_{i_k}=0.$$    $$\sum_{k=2}^{2m+1}s_{i_k}=0.$$ As a result $s_{i_1}+s_{i_{2m+1}}=0.$
So $$\sum_{k=1}^{2m-2}s_{i_k}=0.$$
 Subsequently the sum of $2m-4,\cdots,2$ numbers is 0.
So by Property 2, $s=(1,1,\cdots,1)$.

\noindent{\bf Appendix B. Application for solving the 3SAT problem}\\

\noindent{\bf Lemma 3.}
The 3SAT satisfiable problem can be transformed to determine whether there exists a solution of the equations
\begin{eqnarray}
 (s_{i_1}+r_{i_1})(s_{i_2}+r_{i_2})(s_{i_3}+r_{i_3})=0
 \end{eqnarray}
where $r_{i_j}\in \{0,1\}$ are constants.

\noindent{\bf Theorem 4.} (1)If $ \exists a_{i_1\cdots i_k}=1(k\geq4).
\,\,\forall a_{i_1\cdots i_ki_{k+1}\cdots i_{k+l}}=0(1\leq l\leq r-k).
\,\,a_{i_1\cdots i_{k-4}i_{k-3}i_ji_l}=1(k-2\leq j\leq l\leq k).\,\,
a_{i_1\cdots i_{k-4}i_{k-3}i_j}=1(j=k-2,k-1,k).\,\,
a_{i_1\cdots i_{k-4}i_{k-3}}=1.$
The other coefficients like $a_{i_1\cdots i_{k-4}l_1\cdots l_t}(1\leq t\leq 4)$ are all 0.
Then
\begin{eqnarray}
 &&s_{i_{k-3}}+s_{i_{k-3}}s_{i_{k-2}}+s_{i_{k-3}}s_{i_{k-1}}+s_{i_{k-3}}s_{i_{k}}\nonumber\\
 &+&s_{i_{k-3}}s_{i_{k-2}}s_{i_{k-1}}+s_{i_{k-3}}s_{i_{k-2}}s_{i_{k}}+s_{i_{k-3}}s_{i_{k-1}}s_{i_{k}}\nonumber\\
 &+&s_{i_{k-3}}s_{i_{k-2}}s_{i_{k-1}}s_{i_{k}}\nonumber\\
 &=&s_{i_{k-3}}(s_{i_{k-2}}+1)(s_{i_{k-1}}+1)(s_{i_{k}}+1)=0
 \end{eqnarray}
(2) If $ \exists a_{i_1\cdots i_k}=1(k\geq3).$
  $\forall a_{i_1\cdots i_ki_{k+1}\cdots i_{k+l}}=0(1\leq l\leq r-k).$
 \\(a) If $a_{i_1\cdots i_{k-3}i_{k-2}i_{k-1}}=1,$  $a_{i_1\cdots i_{k-3}i_{k-2}i_{k}}=1,$
$a_{i_1\cdots i_{k-3}i_{k-2}}=1,$
the other coefficients like $a_{i_1\cdots i_{k-3}l_1\cdots l_t}(1\leq t\leq 3)$ are all 0.
Then
\begin{eqnarray}
&&s_{i_{k-2}}s_{i_{k-1}}s_{i_{k}}+s_{i_{k-2}}s_{i_{k-1}}+s_{i_{k-2}}s_{i_{k}}+s_{i_{k-2}}\nonumber\\
&=&s_{i_{k-2}}(s_{i_{k-1}}+1)(s_{i_{k}}+1)=0
 \end{eqnarray}
 (b) If $a_{i_1\cdots i_{k-3}i_{k-2}i_{k-1}}=1,$
the other coefficients like $a_{i_1\cdots i_{k-3}l_1\cdots l_t}(1\leq t\leq 3)$ are all 0.
Then
\begin{eqnarray}
&&s_{i_{k-2}}s_{i_{k-1}}s_{i_{k}}+s_{i_{k-2}}s_{i_{k-1}}\nonumber\\
&=&s_{i_{k-2}}s_{i_{k-1}}(s_{i_{k}}+1)=0
 \end{eqnarray}
(c) If
the other coefficients like $a_{i_1\cdots i_{k-3}l_1\cdots l_t}(1\leq t\leq 3)$ are all 0.
Then
\begin{eqnarray}
s_{i_{k-2}}s_{i_{k-1}}s_{i_{k}}=0
 \end{eqnarray}


\begin{thebibliography}{00}
\bibitem{DR97}
D. R. Simon, On the Power of Quantum Computation, SIAM J. Comp., 26 (1997), pp.1474-1483.
%\bibitem{AAS09}
%A. W. Harrow, A. Hassidim, and S. Lloyd, Quantum Algorithm for Linear Systems of Equations, \emph{Physical Review letters}, 103, 150502 (2009).
\bibitem{SD01}
S. Dubuc, Characterization of Linear Structures, Designs, \emph{Codes and Cryptography}, 22 (2001), pp.33-45.
\bibitem{EC97}
E. Dawson and C. K. Wu, On The Linear Structure of Symmetric Boolean Functions, \emph{Australasian Journal of  Combinatorics}, 16 (1997), pp.239-243.
\bibitem{FX95}
D. G. Feng and G. Z. Xiao , Character of Linear Structure of Boolean Functions,
 \emph{Journal of Electronics} (in chinese), 17(3), 1995, pp.324-329.
\bibitem{CZX11}
D. Z. Cheng, Y. Zhao, and X. R. Xu, From Boolean Algebra to Boolean Calculus, \emph{Control Theory  Applications} (in chinese), 28(10),2011, pp.1513-1523.
\bibitem{PW97}
P. W. Shor, polynomial-time Algorithm for Prime Factorization and Discrete logarithms on Quantum Computer, SIAM J. Comp., 26 (1997), pp.1484-1509.
\bibitem{JY91}
J. Y. Shao, \emph{Combinatorics}(in chinese), Tongji University Press, 1991, pp:86-98.\\
\end{thebibliography}
\end{document}